# Periodic Trends in Adsorption Energies Around Single-Atom Alloy Active Sites

*Julia Schumann*[*,†,§,‡], *Yutian Bao*[†], *Ryan T. Hannagan*[∥], *E. Charles H. Sykes*[∥], *Michail Stamatakis*[§], *Angelos Michaelides**[†,‡]

AUTHOR ADDRESS. [†] Department of Physics and Astronomy, University College London, Gower Street, London WC1E 6BT, UK. [§]Department of Chemical Engineering, University College London, Roberts Building, Torrington Place, London WC1E 7JE, UK. [‡]Yusuf Hamied Department of Chemistry, University of Cambridge, Lensfield Road, CB2 1EW Cambridge, UK. [∥]Department of Chemistry, Tufts University, 62 Talbot Avenue, Medford, Massachusetts 02155, USA.

AUTHOR INFORMATION

**Corresponding Author**

* Angelos Michaelides, E-mail: am452@cam.ac.uk,

* Julia Schumann, E-mail: j.schumann@ucl.ac.uk

ABSTRACT Single-Atom Alloys (SAAs) are a special class of alloy surface catalysts that offer well defined, isolated active sites in a more inert metal host. The dopant sites are generally assumed to have little or no influence on the properties of the host metal, and transport of chemical reactants and products to and from the dopant sites is generally assumed to be facile. Here, by performing density functional theory calculations and surface science experiments, we identify a new physical effect on SAA surfaces, whereby adsorption is destabilised by up to 300 meV on host sites within the perimeter of the reactive dopant site. We identify periodic trends for this behaviour, and demonstrate a zone of exclusion around the reactive sites for a range of adsorbates and combinations of host and dopant metals. Experiments confirm an increased barrier for CO diffusion towards the dopant on a RhCu SAA. This effect offers new possibilities for understanding and designing active sites with tunable energetic landscapes surrounding them.

**TOC GRAPHICS**

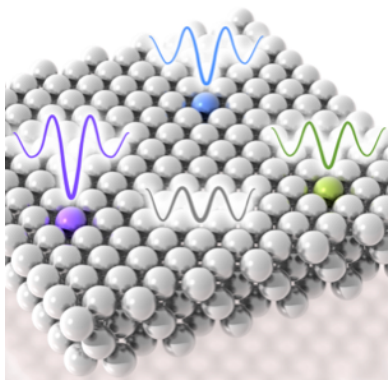

Single-atom alloys (SAAs) are a promising class of transition metal catalysts with well-defined active sites. They have attracted increasing interest in the catalysis community and seen rapid growth of research output in the past decade.[1] SAAs consist of a reactive dopant metal atomically dispersed in a host metal surface.[2] Compared to conventional alloys, the complexity regarding the active site is very much reduced, because of their well-defined structure and the exclusion of ensemble size effects.[1] Additionally, many SAAs display unique electronic structure features as there is limited mixing between the d-states of dopant and host.[3,4] SAAs are especially promising for reactions that conventionally suffer from poor selectivities on the reactive metals or low activity on the more selective coinage metals.[1,5,6] One example is hydrogenation reactions, where the dopant atoms facilitate $H_2$ dissociation, and the H atoms produced can spill over to the host phase, where hydrogenation can occur selectively.[5,7] Hence, the properties of the more selective hosts are favorably combined with highly active dopant atoms.[1,5] SAAs are equally promising for dehydrogenation reactions, where traditional catalysts based on reactive Pt-group metals usually suffer from coking because of the strong adsorption energies of C atoms, whereas the single atom dopants remain coke free.[8] Because of their reactive dopant site, SAAs are particularly promising for applications at low temperatures where reaction selectivity can be maximized. At such temperatures, the rates of adsorbate diffusion across the surface, in order to get to and away from the active site, can be comparable to reaction rates, since these SAA surfaces exhibit low reaction barriers. Diffusion processes towards and away from the dispersed, dopant active sites can therefore become relevant for the overall reaction kinetics. Hence, understanding the energy landscape in the vicinity of the active site is critical for describing its effect on catalysis.

Understanding and predicting adsorption energies on alloy surfaces is a key factor in improving and designing new and better catalysts and recent work builds upon a long tradition of experimental and theoretical work observing and understanding catalytic behaviour and trends on transition metal alloys.[9–13] Usually either geometric (ensemble) effects or ligand effects are cited to contribute to modification of adsorbate binding on alloy surfaces in general.[11,12] How exactly these concepts apply in the very dilute SAA limit is less well studied. The ensemble effect can be considered the dominating factor responsible for their superior catalytic properties in C-H activation.[8] However, the role of the ligand effect is less clear, since limited mixing of d-state between dopant and host matrix has been reported.[3,4]

There are many recent studies about SAAs which report adsorption energies and reactivity at the active dopant site (see e.g. refs. [14–17]), with most studies on SAAs report adsorption energies on the most stable site, which is usually the dopant site.[8,14,15] Different sites, in the vicinity of the dopant on the host surface of the SAA, are rarely considered and compared. Exceptions are works by Hu and co-workers who studied oxygen adsorption on different sites of dilute Pt-based alloys,[18] from which formulas to predict adsorption energies at different sites were developed.[18,19] Recent machine learning approaches by Reuter and co-workers[20] are also able to predict adsorption energies on various sites on a wide range of alloys, including SAAs. However, understanding of the fundamental physical properties responsible for adsorption behavior on different alloy sites is still incomplete.

In this study, we used density functional theory (DFT) to calculate the adsorption of a range of atomic and molecular adsorbates on 33 different SAA surfaces. By comparing adsorption energies at different sites on the alloy surfaces we show that the nearest neighbor sites (referred to as n-sites) that form a ring around the active dopant atom (a-site) have different adsorption properties from the pure host metal surface sites. Specifically, we find that adsorbates bind more weakly to the n-sites than they do on the pure metal host yielding a "zone of exclusion" around the dopant. In addition, we find that, in general, the stronger the adsorption energy on the dopant, the weaker the adsorption energy on the surrounding host metal n-sites. Furthermore, we present experimental evidence that the ring of decreased adsorption energies around the dopant atom is effectively a deterrent that impedes facile diffusion of adsorbates from the host to the dopant. Indeed, kinetic Monte Carlo (KMC) simulations reveal the effect that this zone of exclusion has on the diffusion process of adsorbates from the host to the active site.

In Figure 1a) we illustrate the different a- and n-sites for oxygen adsorption on a model SAA surface. The blue sphere represents the dopant atom and the golden and brown spheres represent the host metal atoms and the n-host atoms surrounding the dopant, respectively. Adsorption of O, and other atomic adsorbates (H, C, N), as well as NO generally occurs preferentially on threefold hollow sites. For consistency throughout this study, we only consider fcc sites as hollow sites, as the fcc and hcp sites have similar adsorption energies on the pure coinage metal hosts. $H_2O$, $NH_3$ and CO molecules are most stable on the atop position and hence placed on atop sites throughout this study.

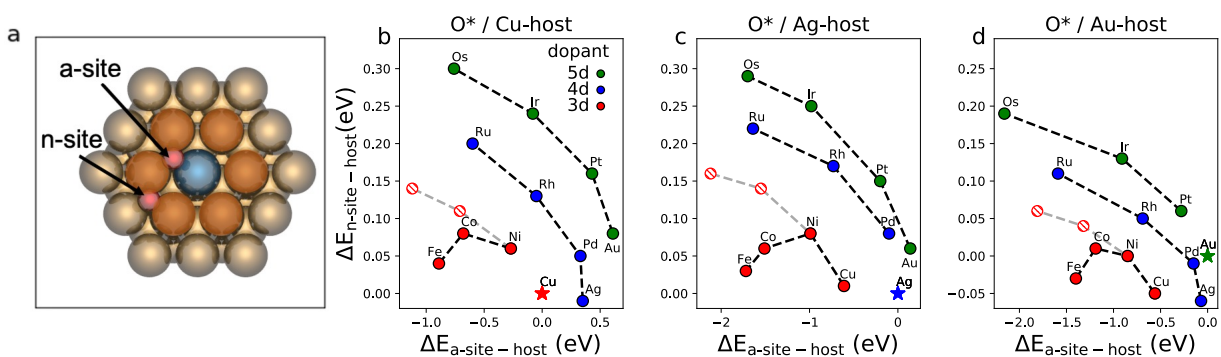

Figure 1. Relative oxygen atom adsorption energies on different single-atom alloy (SAA) sites follow periodic trends. a) Model of the SAA surface indicating dopant a-site and nearest neighbor host n-site for oxygen adsorption. Dopant atom, host atoms, n-host atoms and oxygen

are blue, golden, orange, and red respectively. b-d) Relative oxygen adsorption energies on a-site vs. n-site. Energies are relative to adsorption energy on the respective pure host surface indicated by a star, with positive values meaning weaker adsorption energies. b) Cu-based, c) Ag-based and d) Au-based SAAs. The different points are grouped according to the position of the dopant in the periodic table, 3d-dopants (Cu, Ni, Co, Fe) in red, 4d dopants (Ag, Pd, Rh, Ru) in blue and 5d dopants (Au, Pt, Ir, Os) in green. Stars represent the pure host surfaces. Shaded red points connected with grey lines represent spin-restricted calculations of 3d dopants.

*A zone of exclusion and periodic trends in O adsorption on Cu, Ag, and Au-based SAAs*

We start by showing the results of oxygen adsorption energies, because oxygen is a common adsorbate in catalysis and illustrates the trends very clearly. In Figure 1b-d) we show relative oxygen adsorption energies at the a-site vs. at the n-site, with all energies given relative to the adsorption energy on the pure host surface. We observe that the stronger the adsorption on the dopant site, the weaker the adsorption energy on the neighboring site, within each row of the periodic table. An exception to the monotonic trend is observed for the 3d element dopants. Upon further analysis, we identified the effect of magnetization to be sufficient to "wash out" the zone of exclusion. Removing the effect of spin by doing spin-restricted DFT calculations for Co and Fe, we can see that the same trend as for the 4d and 5d dopants is restored for SAAs with 3d dopants: with increasing O adsorption energy at the a-site, the surrounding n-sites bind oxygen less strongly (see shaded points and grey lines in Figure 1b-d).

At first sight, the trends observed here seem counterintuitive, as we often think about achieving intermediate adsorption energy strengths upon alloying two metals. For example, when using the Sabatier principle to predict new catalysts, it was suggested, that combining a metal with a weak adsorption energy with a strong binding metal would lead to a surface with intermediate binding strength.[21-23] However, in the present case we reduce the weak adsorption energy of the coinage metal even further in the proximity of a more reactive metal atom. Our observation agrees with earlier work from Wang and Hu where they also observe a weakened oxygen adsorption energy at a certain distance to a more reactive metal alloyed in a Pt surface[18].

*Periodic trends for an extended set of adsorbates on Cu-based SAAs*

Next, we explore if the effect observed for oxygen adsorption on the set of SAAs considered, is generalizable to other adsorbates. For this purpose, we show the adsorption energies of a range of atomic and molecular adsorbates on the different n- and a-sites focusing on a range of Cu-based SAAs shown in Figure 2. As we have seen previously for O adsorption, we observe a trend towards a weaker adsorption energy for binding on the n-site, the stronger the adsorption is on the a-site and the further the dopant is to the left of Cu in the periodic table. The 3d dopants Fe and Co again deviate from the monotonic trend due to spin effects, as can be seen through the comparison with spin-restricted DFT calculations indicated in shaded red/light grey in the upper panel for Figure 2.

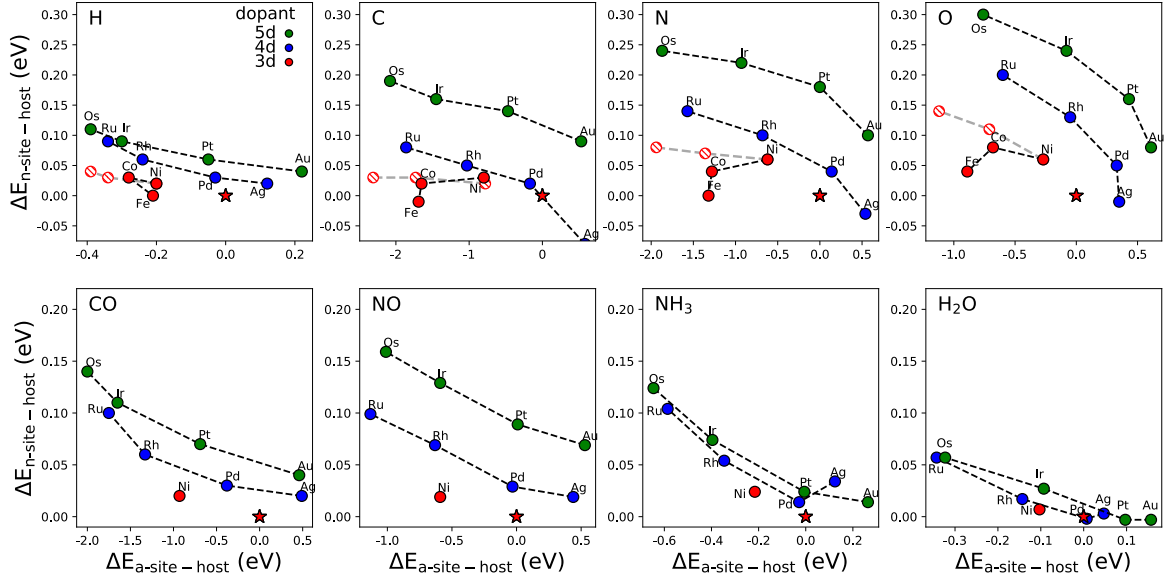

Figure 2. Relative adsorption energies for a range of adsorbates on Cu-based SAAs. Upper panel: H, C, N and O on the threefold n- and a-sites. Lower panel: molecular adsorption of CO, $NH_3$ and $H_2O$ on atop n- and a-sites, and NO adsorption on threefold n- and a-sites. All adsorption energies are given relative to adsorption of the respective adsorbate on pure Cu. Weakened adsorption on the n-site (y-axis, mostly positive values), stronger adsorption on the a-site (x-axis, mostly negative energies). Green, blue and red colors for 5d, 4d, and 3d dopants, respectively.

The monotonic trend for dopants in the same row of the periodic table, leading to a weakened adsorption energy surrounding the dopant, is furthermore not only valid for atomic adsorbates, but also for closed and open shell molecules as shown in the lower panel of Figure 2. For strongly bound adsorbates such as CO and NO, and likewise for more weakly bonded adsorbates such as $NH_3$ and $H_2O$, all of them show a general trend with weaker adsorption around the perimeter of the dopant compared to pure Cu. Exceptions to the weakened binding on the n-sites are some adsorbates on Ag-doped systems, for which adsorption at the a-site is even weaker, as well as Fe-based systems due to spin effects. We generally observe that for adsorbates such as H, $NH_3$ and $H_2O$ where the adsorption energy is relatively weak, the range of adsorption energies at the a-site of different surfaces is small (as seen by the small range of the x-axes in Figure 2). At the same time, for these weakly binding adsorbates the differences in adsorption energies on the n-site and the separation of the trend lines of the 3d, 4d and 5d dopants are also smaller. We also note that for $H_2O$ adsorption on the PtCu and PdCu SAAs, where the adsorption on the dopant itself is weak, the adsorption along the perimeter atoms is hardly changed compared to pure Cu. For strongly adsorbing species on the other hand, with a range of adsorption energies on the a-site larger than 1 eV, the effect of weakened adsorption strength on the n-site is larger.

*Experimental evidence for a zone of exclusion*

Thus far, we have presented computational evidence for a weakened adsorption energy surrounding the dopant in SAA surfaces. Now, we will discuss experimental observations that corroborate these computational results. Using a combination of temperature programmed desorption (TPD) and reflection absorption infrared spectroscopy (RAIRS), we observed that diffusion of CO from the Cu(111) host to Rh dopant atom sites is an activated process with an increased diffusion barrier to what one would expect on pure Cu(111). This effect is consistent with a ring of decreased CO adsorption energy around the reactive dopant, which effectively acts as an additional diffusion barrier and impedes CO diffusion towards the dopant Rh atom, in agreement with our computational results.

Figure 3 shows a summary of these experimental results comparing a clean Cu(111) with a RhCu(111) SAA surface. A detailed characterization of model RhCu SAA surfaces has been reported previously[24,25] and more data supporting the successful formation of the SAA is shown in Fig. S10 in the SI. The TPD spectra of the two surfaces in Figure 3a confirm that a strongly adsorbed CO species exists on the RhCu(111) SAA. Specifically, the black TPD trace shows that CO binds weakly to Cu(111) and desorbs from the surface at <220 K. When 0.2% Rh is alloyed into the Cu(111) surface the CO desorption traces look almost identical to the clean Cu(111) case except for a new desorption peak at ~470 K corresponding to CO desorption from isolated Rh atoms which bind CO much more strongly, in agreement with our DFT calculations. Usually, it is assumed, that the strongly adsorbing sites are the first sites to become occupied when adsorbates like CO are deposited on the surface. However, as shown by the CO-RAIRS experiments, which are sensitive to the binding site of the adsorbed CO (Figure 3b and 3c), we found that after depositing CO onto the RhCu(111) SAA surface at 90 K under ultra-high vacuum (UHV) conditions, no CO was detected on the Rh atom sites and IR peaks around 2100 $cm^{-1}$ characteristic of CO on Cu(111) were observed. Only after heating the surface to 117 K did we observe a growing population of CO on isolated Rh atop-sites, as evidenced by the appearance of an IR peak at ~2000 $cm^{-1}$ characteristic of CO on isolated Rh sites.[24] As CO binds very weakly on Cu(111), a high mobility of CO can be assumed at 90 K,[26] which should lead to a fast equilibration of CO on the surface and population of the Rh-sites. However, the irreversible temperature dependent population of CO on the strong binding Rh-site indicates that the process is activated and provides evidence for an increased diffusion barrier surrounding the Rh atoms.

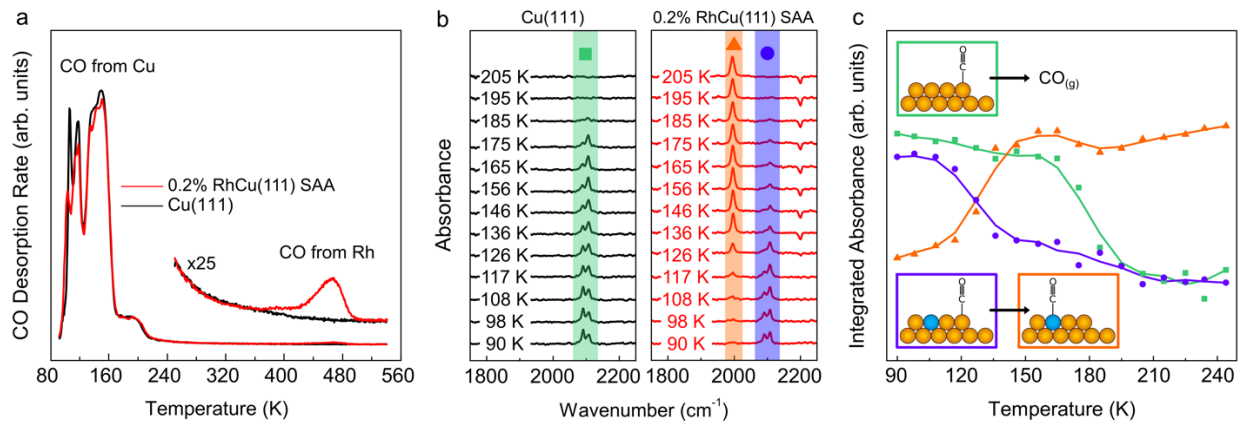

Figure 3. Surface science experiments showing impeded diffusion of CO from the Cu(111) host to the Rh atom dopants in the RhCu(111) SAA. a) TPD spectra of Cu(111) (black line) and 0.2% RhCu(111) SAA (red line) after dosing a saturated monolayer of CO at 90 K. Inset red trace shows a small fraction of CO desorption from RhCu(111) SAA at elevated temperature b) RAIRS spectra of Cu(111) (left) and RhCu(111) SAA (right) collected after dosing ~2% monolayer CO at 90 K and heating in increments of ~10 K from 90 to 205 K (see SI for details and full temperature range up to 244 K). c) Integrated IR peak areas of CO adsorbed on pure Cu(111) (green), Cu sites of the RhCu(111) SAA (purple), and Rh atom sites of the RhCu(111) SAA (orange) as a function of temperature. Solid lines are added as guides to the eye.

*KMC reveals a strong influence of the zone of exclusion on diffusion*

To assess the effect such a ring of weaker binding sites around the single atom dopant has on the diffusion process, we performed KMC simulations on a model SAA lattice. To see the effect of a destabilized n-site around a strongly binding active site we measured how fast CO moves to the dopant site. To this end, we determined mean first passage times of a randomly placed CO molecule on a RhCu SAA surface compared to a SAA surface without a decreased adsorption energy at the n-sites surrounding the dopant (Figure 4a). Our simulations show that the mean first passage times of the CO adsorbate to reach the Rh dopant from the Cu host differ by three orders of magnitude at 100 K, when introducing a plateau of decreased adsorption energy at the n-site. By introducing a "ring" of neighboring sites with lower adsorption strength around the dopant, even with all the forward activation barriers being constant, CO diffusion towards the dopant is hindered. This is in good agreement with the experimental data, which suggests that, below 117 K, the diffusion from the Cu to the Rh dopant is too slow to be observed at the timescale of the experiment. KMC simulations indeed confirm a delay of CO diffusion to the Rh atom site by more than three orders of magnitude at 100 K (see Figure 4a). It can be seen from the section of a random walk of CO on the KMC lattice in Figure 4b, that this delay does not stem from a lack of opportunities for CO to reach the dopant site as there are plenty of occasions where the CO is near to or next to the dopant. This pictorial representation is confirmed by the plot of distances between the adsorbate and its nearest a-site, shown in Figure 4c. Clearly the destabilization (of 0.07 eV) at the n-site leads to a preferential diffusion away from the dopant to the host in marked contrast to the behavior seen in the absence of the destabilizing n-sites (c.f. blue traces in Figures 4c and d).

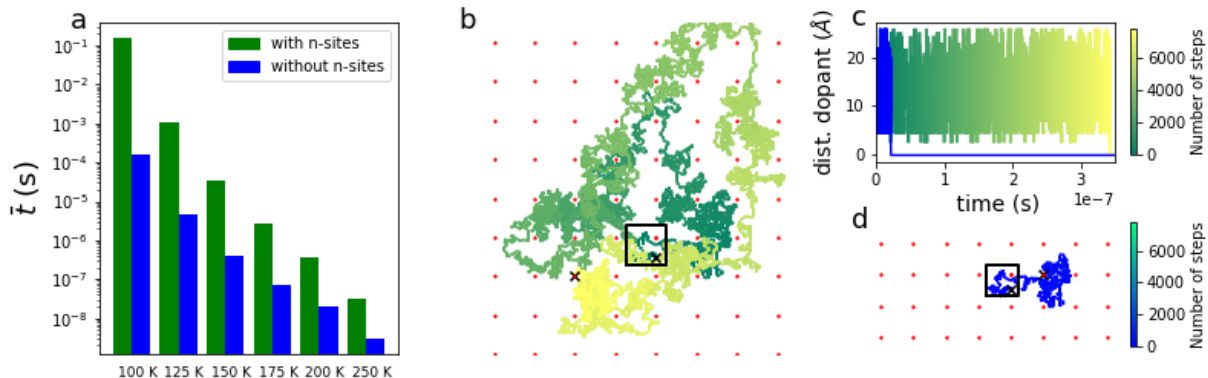

Figure 4. KMC simulations show that the zone of exclusion significantly increases the time CO molecules take to reach the dopant site in a RhCu SAA. a) Mean first passage times $\bar{t}$ of CO reaching the dopant site at 6 different temperatures between 100 K and 250 K, determined from 200 KMC runs with different random seeds at each temperature, on SAAs with and without specification of n-sites in the lattice. b) Example of a random walk of a CO molecule on a RhCu SAA surface at 200 K. The unit cell of the periodic KMC lattice with a size of 120 top sites (black box) and the position of dopant sites in each unit cell (grid of red points) are shown. CO reaches the dopant site for the first time after >7,000 diffusion steps in this representative KMC run. Start and end points are marked with black crosses. c) Distance of CO adsorbate to dopant site as a function of time. d) Example of a random walk of a CO molecule on a RhCu surface without n-sites surrounding the dopant.

Having identified a zone of exclusion that is present in the vicinity of reactive dopants in SAAs, we now briefly comment on the physical origin of the effect. Several excellent models have been developed to predict or rationalize adsorption energy variations on transition metal alloy surfaces (see e.g. refs [18,19,27–32]). Two main factors are usually relevant, namely geometric (strain and ensemble)[33–35] and electronic effects,[13,34,36,37] with the d-band center of the alloy surface often reflecting how these properties change.[32,36,37] Indeed, some of the trends observed here on the Cu-based SAA can in principle be rationalized with strain effects. Specifically, the larger dopants dispersed in a smaller Cu lattice lead to compressive strain of the surrounding Cu host atoms and result in weaker adsorption on the n-site. This can also explain the offset between the 4d and 5d dopants (see Fig. S9a in SI). However, strain does not adequately account for the trends observed on the Ag- and some of the Au-based SAAs. Furthermore, no significant d-band center shifts of the n-sites relative to the pure host are observed, as one would expect for a uniformly strained host (see Fig. S7 in SI). Indeed, careful analysis of structures and bond lengths in the vicinity of the dopant show that changes in the host-host distances are non-uniform with certain bonds getting compressed and others stretched. Thus, it is not a surprise that prominent shifts in the d-band center are not observed.

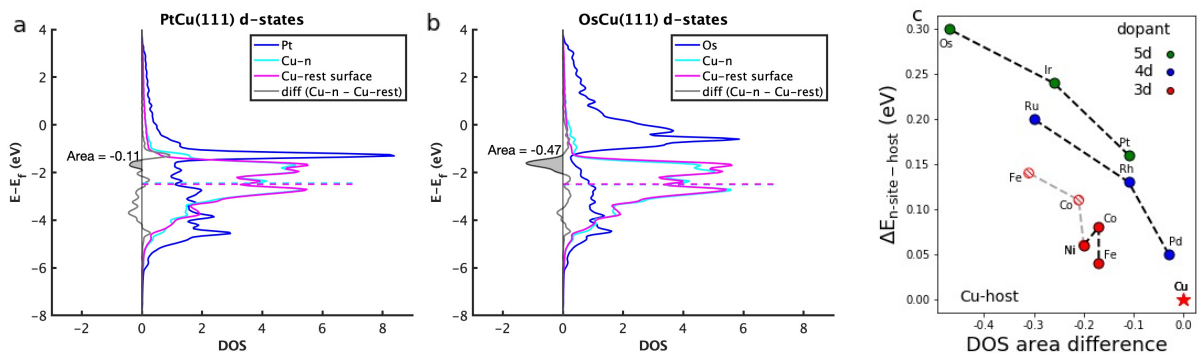

Figure 5. Density of states (DOS) analysis of clean SAAs. a-b) d-state DOS of surface atoms of (a) PtCu-SAA and (b) OsCu-SAA. The d-states of the single atom dopant are indicated in blue, of the n-site Cu atoms in cyan, and of the rest of the Cu surface atoms (2nd and 3rd shell around dopant) in magenta. The differences between n-site Cu atoms and the remaining Cu surface atoms are indicated in grey. The shaded grey area corresponds to the difference in d-state DOS at the upper band edge of the Cu atoms. Dotted lines indicate the position of d-band center of the Cu-surface atoms. c) DOS area difference of upper d-band edge of the n-Cu atoms and the rest of the surface Cu-atoms versus weakened oxygen adsorption strength on the n-site.

What then can explain the observed weakened adsorption energies at the n-sites? Detailed analysis of the d-band density of states (DOS) reveals a decrease of the DOS at the upper band edge of the n-Cu atoms compared to the rest of the Cu surface atoms (Fig. 5a-b and Fig. S6). In addition, the magnitude of this reduction correlates with the periodic trends observed (Fig. 5c).

This reduction in states at the upper d-band edge leaves fewer of the relatively high-lying d states available for bonding with adsorbates. Essentially, this analysis reveals an intuitive explanation, regarding the stability of the metal site itself: the stronger a metal atom is bound to the rest of the surface, the smaller is its capacity to bind to an adsorbate.[38] A host metal atom that has a reactive dopant neighbor, can be expected to experience a strong bond with this dopant. Also, the larger the covalent d-band contribution of the dopant, the stronger the effect on the sourrounding host matrix. This compensating effect has not only been observed for differently coordinated pure metal surfaces,[35,38] but also for Pt-based alloys,[18] where similar effects of decreased oxygen adsorption energies at $2^{nd}$ nearest neighbor positions around dopants have been seen and formulas to predict adsorption energies on the different sites were found empirically.[19] These observed trends of adsorption energies also agree with many conventional chemical trends of the periodic table, such as cohesive energies. Thus, overall, we find that a combination of strain and electronic effects of the SAA surface leads to bond compensation effects, which ultimately result in a weakened adsorption energy on the n-sites around the dopant.

Before concluding, we note that, as well as slowing down diffusion towards the dopant at low temperatures, the zone of exclusion around the dopant could influence catalytic processes through confinement effects leading to longer residence times at the dopant site. Confinement of two adsorbates which are either reactants or products of a catalytic reaction, could lead to different product selectvities in cases where several reaction pathways are possible, such as in dehydrogenation reactions of longer chain alkyls. Furthermore, cases where previously predicted behaviour did not match experimental observations will have to be evaluated anew to determine whether exclusion or confinement effects would be able to correct for previous mismatches. Whether our predicted effects will hold true in a nanoparticle catalyst under reaction conditions, we cannot say at this point. However, our recent results[6] of a successfully designed nanoparticle catalyst from theoretical predictions are showing the promising capability of our approach.

Apart from direct consequences for catalysis, we could imagine that the zone of exclusion could also be useful for creating controlled facile diffusion paths on metal surfaces along certain directions, without the adsorbate necessarily being trapped on strong binding sites that may guide the diffusion path at low temperatures.[39]

In conclusion, on the basis of a broad DFT screening, a novel general effect for SAAs is predicted where a reactive dopant site induces a weakening of adsorption strength on the surrounding host metal sites. Detailed experimental measurements on RhCu SAAs surfaces at low temperatures have demonstrated that this effect manifests itself as an increased diffusion barrier around the dopant, leading to a delay of CO diffusion from the host Cu surface to the Rh atom dopant. This so-called zone of exclusion has implications for understanding and designing active sites for reactions run at low temperatures with the aim of achieving high product selectivity.

Methods section – All DFT calculations were performed using the Vienna *Ab Initio* Simulation Package (VASP) version 5.4.4[40,41] with the projector augmented wave (PAW) method to model core ionic potentials and the non-local optB86b-vdW exchange-correlation functional[42,43]. The surfaces were modeled using a 4-layer p(5×5) slab separated by a 15 Å vacuum layer, and a 7×7×1 Monkhorst-Pack k-point mesh was chosen for Brillouin-zone integration. The relative adsorption energies $\Delta E_{n/a-site\ -\ host}$ used in this study were calculated as follows:

$$\Delta E_{n/a-site\,-\,host} = E - E_{SAA\,clean} - (E_{host} - E_{host\,clean}),$$

with E being the DFT total energy of the SAA slab with adsorbate, $E_{SAA\,clean}$ the DFT total energy of the clean SAA slab, $E_{host}$ the DFT total energy of a pure host surface with an adsorbate, and $E_{host\,clean}$ the DFT total energy of the clean pure host metal slab. Convergence tests and further computational details can be found in the SI.

KMC model simulations were performed using the graph-theoretical KMC framework as implemented in Zacros version 2.0.[44–46] The simulation cell was formed of (6 × 10) rectangular unit cells of an fcc(111) surface, each containing two atop sites, resulting in 120 atop sites per simulation cell. Periodic boundary conditions were applied. Simulations for the Rh/Cu(111) SAA made use of a lattice in which one Cu site was randomly substituted with a Rh site, resulting in a density of 0.8% Rh/Cu. The Rh site was surrounded by 6 atop n-sites and 6 threefold n-sites. Given the known difficulty of determining CO adsorption energies and site preference correctly with DFT,[47–50] adsorption energies for CO were estimated from DFT[51] for Cu atop sites and set to the values of -0.80, -0.73 and -2.04 eV on host sites, n-sites, and a-site respectively, based on the energy differences determined from DFT. Forward diffusion barriers (from host to n-site to a-site) were set to 0.15 eV, based on estimates linking adsorption energies with respective diffusion barriers from previous work.[52] The first passage time of a randomly placed CO adsorbate moving to the dopant was determined and the mean was calculated from an ensemble of 200 random seeds for each different temperature between 125 and 250 K.

TPD and RAIRS experiments were conducted in a home-built UHV chamber. Details of the chamber characteristics, equipment, and sample preparation can be found in the SI.

ASSOCIATED CONTENT

**Supporting Information**.

The following files are available free of charge.

Detailed information about the computational and experimental setup, convergence tests and further computational and experimental data can be found in the SI. (PDF)

Optimized structure files are included in a zip file. (POSCAR)

AUTHOR INFORMATION

**Notes**

The authors declare no competing financial interests.

ACKNOWLEDGMENT

Via our membership of the UK's HEC Materials Chemistry Consortium, which is funded by EPSRC (EP/L000202, EP/R029431), this work used the UK Materials and Molecular Modelling Hub for computational resources, MMM Hub, which is partially funded by EPSRC (EP/P020194 and EP/T022213/1). M.S. and A.M. acknowledge funding from the Leverhulme Trust, grant ref. RPG-2018-209. Furthermore, we thank UCL Research Computing Services for the use of Grace, Kathleen and Myriad. J.S. is grateful for support through a Feodor Lynen Fellowship of the Alexander von Humboldt Foundation. The experimental part of the work (R.T.H. and E.C.H.S.) was funded by the Division of Chemical Science, Office of Basic Energy Science, CPIMS Program, U.S. Department of Energy, under Grant DE-SC 0004738.